\documentclass[iop,apjl]{emulateapj}
\usepackage{amssymb}
\usepackage{amsmath}
\usepackage{hyperref}
\usepackage{natbib}
\usepackage{color}
\usepackage{graphicx}

\begin{document}
\title{Strong Preferential Ion Heating is Limited to within the Solar Alfv\'en Surface}

\author{Justin ~C. Kasper$^{1,2}$
Kristopher ~G. Klein$^{1,3}$
}

\affiliation{$^1$Climate and Space Sciences and Engineering, University of Michigan, Ann Arbor, MI 48109, USA
$^2$Smithsonian Astrophysical Observatory, Cambridge, MA 02138 USA
$^3$Lunar and Planetary Laboratory, University of Arizona, Tucson, AZ 85719, USA}

\begin{abstract}
The decay of the solar wind helium to hydrogen temperature ratio due to Coulomb thermalization can be used to measure how far from the Sun strong preferential ion heating occurs.  Previous work has shown that a zone of preferential ion heating, resulting in mass-proportional temperatures, extends about $20-40 R_\odot$ from the Sun on average.  Here we look at the motion of the outer boundary of this zone with time and compare it to other physically meaningful distances.  We report that the boundary moves in lockstep with the Alfv\'en point over the solar cycle, contracting and expanding with solar activity with a correlation coefficient of better than 0.95 and with  an RMS difference of $4.23 R_\odot$.  Strong preferential ion heating apparently is predominatly active below the Alfv\'en point.  To definitively identify the underlying preferential heating mechanisms, it will be necessary to make in situ measurements of the local plasma conditions below the Alfv\'en surface. We predict Parker Solar Probe (PSP) will be the first spacecraft to directly observe this heating in action, but only a couple of years after launch as activity increases, the zone expands, and PSP's perihelion drops. 
\end{abstract}


\maketitle
\section{Introduction}

Ions in the solar corona and solar wind are too hot, a puzzle since the start of the space age and the first in situ observations of solar wind plasma by spacecraft.  Since electrons carry the heat flux and are the ultimate sink of turbulent energy, one would expect electrons to be hotter than ions, and for this difference to grow more extreme with distance from the Sun; however, ions are hotter in the corona and have similar temperatures to electrons at 1 au.  Explaining how ions are heated in the corona and solar wind remains a major challenge in the field.    Spectroscopic observations allow us to remotely observe the onset of this preferential ion heating and its consequences.  A few tenths of a solar radius $R_\odot$ above the Sun's photosphere, rising temperatures and falling densities greatly diminish the Coulomb collisions that enforce thermal equilibrium. Unidentified plasma heating mechanisms in this region couple to different ion species with varying efficiency, preferentially heating some ions more rapidly than others.  Extreme differences in ion temperatures develop, with some species reaching temperatures beyond $100$ MK \citep{Parker:1988,Kohl:1998,Landi:2009}.  The ratio of heavy ion species temperature to proton temperature $T_s/T_p$ is observed to reach and even exceed the mass ratio $m_s/m_p$.  This suggests a kinetic heating process involving interactions with waves or fluctuations with a characteristic velocity, as ions have equal thermal speeds when they have mass proportional temperatures.  Remote observations provide some insight into the mechanisms injecting energy at the base of the corona \citep{McIntosh:2011,Grant:2018} but these observations are not sufficient to distinguish between the various mechanisms that have been proposed to lead to preferential ion heating throughout the near-Sun environment, including wave damping, turbulent dissipation, shocks, reconnection, nano-flares, and velocity filtration (see reviews by \citet{Ofman:2010}, \citet{Hansteen:2012} and \cite{Cranmer:2012}).  
At some distance, strong preferential ion heating ceases, and ion temperature differences in the solar wind begin to decay with increasing time, as infrequent Coulomb collisions begin to thermalize the plasma \citep{Neugebauer:1976,Hernandez:1987,Tracy:2015}.   
We will refer to the heating process active near the Sun that results in mass proportional, and even super-mass proportional temperatures, as strong preferential heating.
In solar wind far from the Sun, mass proportional temperatures are only observed when the frequency of Coulomb collisions is low.  Helium and heavier ions in the solar wind with high Coulomb collision rates are at most tens of percent hotter than protons, which is either an indication of the temperature measurement error of the instrument or a sign that only much weaker preferential ion heating occurs in interplanetary space \citep{Maruca:2013,Tracy:2016}.

Recently we demonstrated a technique for using solar wind observations
at 1 au to determine how far from the Sun the strong preferential
ion heating occurs (\cite{Kasper:2017}, referred to as Paper I).
We proposed that there is a zone
close to the Sun where ion species experience strong preferential heating, and that
within this zone of preferential heating ions reach an equilibrium temperature with an unspecified
heating mechanism resulting in different steady temperature ratios for
different ion species relative to protons.  The start of this zone is seen in
the spectroscopic observations just a few $0.1 R_\odot$ above the
photosphere.  We further assumed that there is an outer boundary of the
zone, at a distance $R_b$ from the Sun, beyond which strong preferential heating ends, and ions are either heated equally or at some much weaker preferential rate.  Beyond $R_b$, Coulomb relaxation, or the accumulated impact of many small angle Coulomb scattering interactions between ions, dominates over any weak preferential heating, and will slowly act to drive species toward equal temperatures.  For an intuitive sense of this process, consider \cite{Spitzer:1962} who showed that if two species have a
difference in temperature $\Delta T$ and exchange thermal energy via Coulomb scattering at frequency $\nu_c$, with only one species having a notable change in temperature, then the temperature difference will change with time as
\begin{equation}
\frac{d\Delta T}{dt}=-\nu_c\Delta T.
\end{equation}
Rearranging and defining the temperature excess $\epsilon\equiv T_\alpha/T_p-1$,  where $T_\alpha$ and $T_p$
are the temperatures of fully ionized helium and hydrogen yields
\begin{equation}
\epsilon(A_c)=\epsilon_o e^{-\int \nu_c dt}=\epsilon_o e^{-A_c}
\label{eqn:eps}
\end{equation}
where $A_c$ is the Coulomb age, or the number of Coulomb thermalization times that have elapsed from when the plasma crossed $R_b$ to when it was observed in space, and $\epsilon_o$ is the steady state excess temperature ratio developed below $R_b$.  Both helium and minor ions seen at 1 au exhibit exponential-like decay in temperature excess with $A_c$\citep{Tracy:2015,Kasper:2017}.  We can locate $R_b$ by using the exponential decay of $\epsilon(A_c)$ as a clock to measure the time it took the solar wind to move from the outer boundary of the zone of preferential heating to the observing spacecraft.  
In this paper, we report for the first time the temporal dependence of $R_b$, and find that this outer boundary of strong preferential ion heating is well correlated with the Alfv\'en critical surface. Implications for in situ observations of preferential heating mechanisms are discussed.

\section{Methodology}
This work uses measurements from the Solar Wind Experiment \citep{Ogilvie:1995} and
Magnetic Field Investigation \citep{Lepping:1995}
instruments on the NASA Wind spacecraft. The same data selection
criterion were used as were presented in Paper I with the added exclusion of data collected
prior to October 27, 1997. Before this date, a different
observation mode yielded larger measurement uncertainties for $T_\alpha$.

Our model for $A_c$ and $R_b$ uses radial power law
exponents $\delta$ and $\sigma$ to capture how overall temperature and
speed vary with distance from the Sun,
$T(r)=T_0 (r/R)^{-\delta}$ and
$U(r)=U_0 (r/R)^{-\sigma}$.  
The solar wind proton density and magnetic field amplitude scale as 
\begin{equation}
n_p(r)=
n_0 R^{2-\sigma} r^{\sigma - 2}
\end{equation}
and
\begin{equation}
B(r)=B_0 R^2r^{-2}\sqrt{1 + \omega^2
  r^{2+2\sigma}U_0^{-2} R^{-2 \sigma}}, 
\end{equation}
where the angular frequency of the Sun's rotation in the equatorial plane is $\omega =
2.7 \times 10^{-6}$ rad s$^{-1}$. {Quantities with a subscript $0$ are
  values measured at $R=1$au}.
  
  Our full model for $\epsilon(A_c)$ is more complicated than Eqn~\ref{eqn:eps} because it allows both species temperatures to change, with the relative rates of heating and cooling determined by the relative mass density $F\equiv m_\alpha n_\alpha/m_p n_p$. An increase in $F$ causes a faster equalization of $T_p$ and $T_\alpha$.
We also account for the variation of $\nu_c$ due to these
relative temperatures changing with distance; see Eqn. 11 in Paper I for the full expression.

In Paper I, we used a range of $\delta$ that bracketed
published values seen by Helios \citep{Marsch:1982,Hellinger:2011} and assumed
$\sigma$ was zero. A best fit for observed $\epsilon$ as a function of
$A_c$, with $R_b$ and $\epsilon_o$ as free parameters was found.  The
analysis in Paper I and in this
work is limited to solar wind intervals where the speed is between
$300$ and $500$ km/s to ensure good data coverage of both high and low collisional age plasma. Using all solar wind speeds does not
qualitatively affect the results presented here.  
Over the entire Wind mission the model fits the
observations with a Pearson's chi-squared test of $\chi^2/$dof of less
than 2, and can predict the mean $\epsilon$ for a given $A_c$ with an
RMS error of less than $10\%$.  If $\delta$ is specified, the best fit uncertainty in $R_b$ is much less than one $R_\odot$.  For every $0.1$ increase in $\delta$, $R_b$ drops $8.8$ $R_\odot$ closer to the Sun. For the range of $\delta$ reported in the literature $R_b$ could be between $20-40 R_\odot$.

\section{Results}

\begin{figure}[ht]
\centering
\includegraphics[height=1.0\linewidth,angle=90,viewport=50 50 530 685, clip=true]
{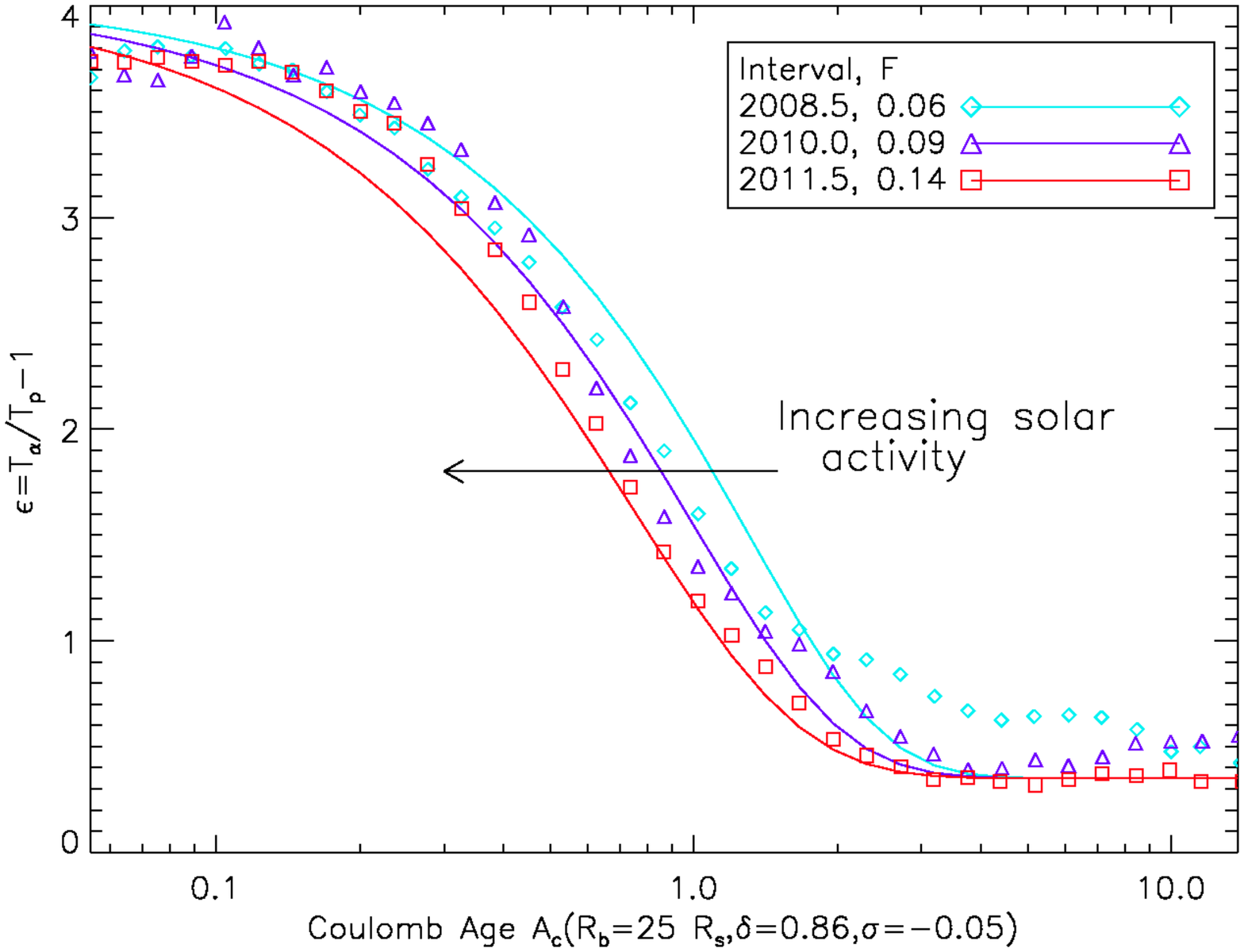}
\caption{
Observed $\epsilon (A_c)$ for three year-and-a-half intervals with increasing solar activity, colored symbols. Solid lines are predicted $\epsilon(A_c)$ for each interval, factoring in the increasing observed mass density ratio $F$ but holding $R_b=25_\odot$.
At solar minimum, $25 R_\odot$ is an overestimate of $R_b$, with the observed $\epsilon$ decaying faster than the model. As activity increased $R_b$ is underestimated, suggesting it is moving outwards from the Sun.}
\label{fig:eps}
\end{figure}

Temporal variation in $\epsilon(A_c)$ and the outer boundary $R_b$ can be seen in a relatively raw view of the observations.  Fig.~\ref{fig:eps} compares the observed $\epsilon(A_c)$ (symbols) with the expected decay (lines) if $R_b$ is taken to be constant but the mass density ratio $F$ is updated to account for its observed solar cycle dependence.  Over these three 1.5-year intervals $F$ grows from $0.06$ to $0.14$ \citep{Kasper:2012}, and with more helium, the two species can reach thermal equilibrium faster.  Most of the observations fall below the curve in solar minimum (light blue diamonds), and are generally all above the curve as activity increases (red squares). This can be explained as an underestimate of $A_c$ in solar minimum because $R_b$ is closer to the Sun than we assumed, and an overestimate of $A_c$ in times of high activity because $R_b$ has moved closer to the observer than assumed. The elevated $\epsilon$ at large $A_c$ in 2008 appears to be to due a higher uncertainty in the temperature of helium caused by lower helium densities which does not impact this analysis.

\begin{figure}[ht]
\centering
\includegraphics[height=1.0\linewidth,angle=90,viewport=15 25 475 665, clip=true]{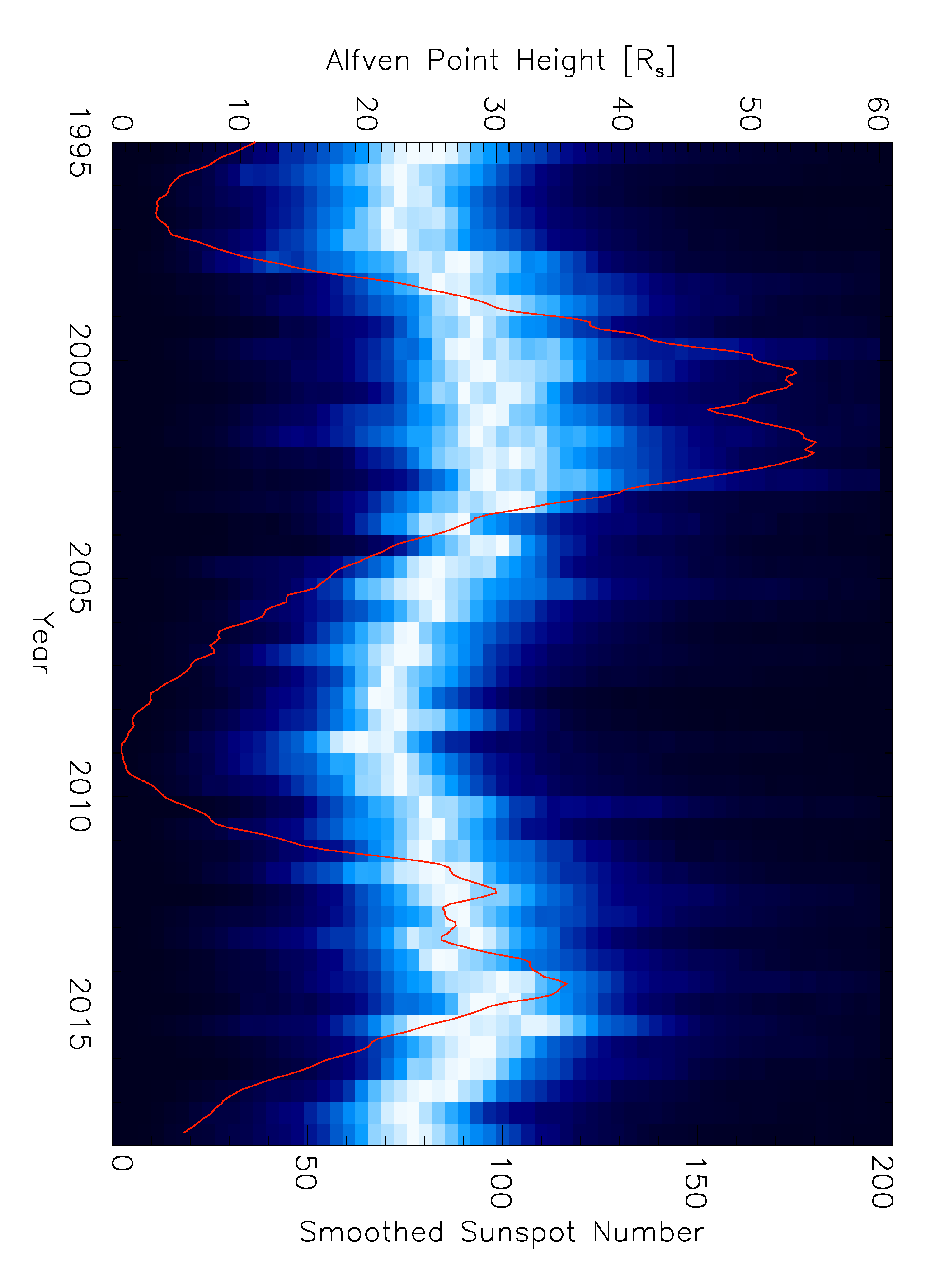}
\caption{A two dimensional histogram of the height of the Alfv\'en
  surface $R_A$ over time calculated using millions of individual
  measurements of the solar wind at 1 au (shades of blue) along with
  the smoothed sunspot number (red line).}
\label{fig:ssn}
\end{figure}

We next calculate a best fit value for $R_b$ as a function of year over the entire Wind mission, 
and compare it to a proxy for solar activity and to several critical surfaces
surrounding the Sun, where the bulk solar wind speed transitions from
below to above some characteristic wave speed.
It is also known that critical surfaces around the Sun also
have temporal dependencies associated with the solar
cycle\citep{Katsikas:2010}. We focus on two of these surfaces,
related to two fundamental waves speeds in a magnetized plasma,
the Alfv\'en speed $v_A$ and the sound speed $v_s$.  For a radial solar
wind profile $U(r)$ the radial location of the Alfv\'en and sound
critical surfaces, $R_A$ and $R_s$, are where $v_A(r)=U(r)$ and
$v_s(r)=U(r)$.  
Beyond these critical surfaces, an Alfv\'en or sound wave
respectively cannot travel back to the Sun.
The Alfv\'en speed is calculated as 
\begin{equation}
    v_A(r) = \frac{B(r)}{\sqrt{4 \pi n_p(r) m_p}}.
\end{equation}
Using an Alfv\'en speed with only the proton mass density or the total ion mass density does not significantly alter the results presented below.
For the sound critical surface $R_s$, we use Eqn 11 from \cite{Katsikas:2010} evaluated at equatorial latitudes.
For each measurement at 1au, assuming a particular radial scaling for the solar wind temperature $(\delta)$ and velocity $(\sigma)$, we determine $R_A$ and $R_s$.
The distribution of $R_A$ as a function of time is shown in Fig.~\ref{fig:ssn} as a column-normalized two-dimensional histogram;
$R_A$ typically has values around $25 R_\odot$ with significant expansion and contraction between solar minimum and solar maximum.

\begin{figure}[ht]
\centering
\includegraphics[width=.95\linewidth,viewport=15 0 290 165, clip=true]
{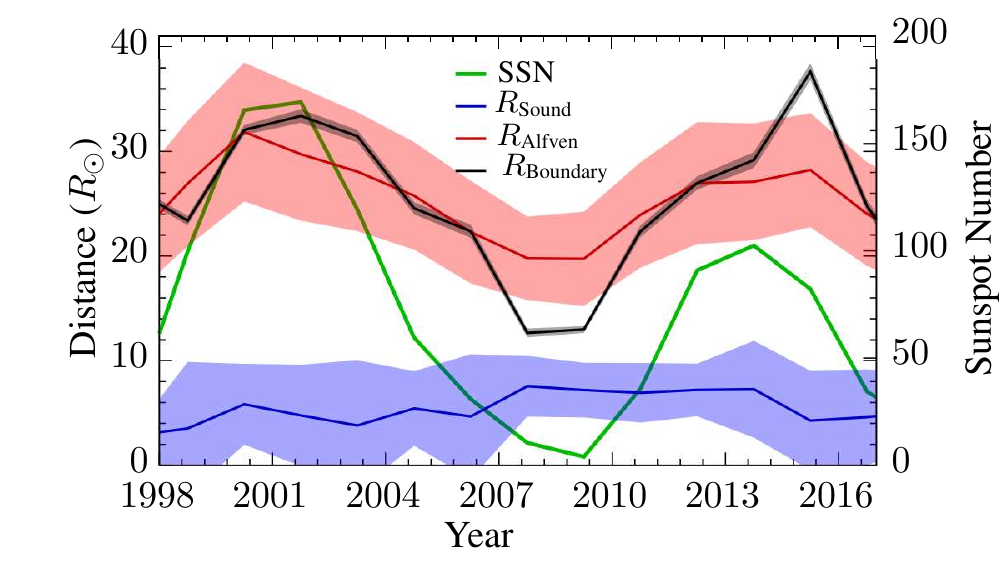}
\caption{Comparison of the motion of the outer boundary $R_b$ of the
  zone of preferential ion heating (black) with the sound surface
  $R_s$ (blue) and Alfv\'en surface $R_A$ (red) as a function of time
  for {$\delta=0.814$ and $\sigma=-0.05$}. The width of the line for $R_b$
  indicates the uncertainty from the model's fit. The red and blue
  shaded regions indicate the RMS variation of $R_s$ and $R_A$ about
  their means.  The yearly averaged Sunspot number (green) is also shown.
  }
\label{fig:distance}
\end{figure}

The outer boundary of the zone of
preferential heating $R_b$ is calculated using the same scheme as in Paper I, 
except instead of segregating by solar wind speed, the data is sub-divided into one-and-a-half year intervals.
The mean value of $R_b$ as a function of time from 1998 to 2017 
is plotted in Fig.~\ref{fig:distance} for values for radial scalings of
proton temperature and speed consistent with previous measurements of the solar wind, ($\delta =0.814$ and
  $\sigma=-0.05$, c.f. \cite{Hellinger:2011,Hellinger:2013}). We observe significant
variation in $R_b$, ranging from $\sim 10 R_\odot$ at solar minimum
to $\sim 35 R_\odot$ at solar maximum.  This variation occurs for
both the relatively strong cycle 23 and the weaker cycle 24.  The
one-sigma error in $R_b$ is quite narrow, on the order of $0.5
R_\odot$.  To compare with $R_b$, we also calculate average values
and RMS variations of $R_A$ and $R_s$, plotted in red and blue in
Fig.~\ref{fig:distance}. The sound critical surface $R_s$ is much
closer to the Sun than $R_b$ and does not have the same temporal
variation with solar activity.  
The Alfv\'en critical surface's temporal variation is well correlated with $R_b$, with 
a RMS difference between the two distances of less than $4.23 R_\odot$ and a Spearman rank
correlation of $0.956$ with significance of $3\times10^{-6}$.
The correlation is
significantly better than with sunspot number, a typical indicator of solar activity $(0.842)$, or the sound
surface $(-0.367)$.

Another physically meaningful distance is where the ratio of thermal and magnetic pressures, $\beta(r)=8 \pi n_p(r) k_B T_p(r) B(r)^{-2}$, 
crosses some critical value;
we denote this surface as $R_\beta$.  Choosing different critical $\beta$
values of less than unity, we find that $R_\beta$ and $R_b$ are well
correlated, with a Spearman rank correlation coefficient of $0.820$,
though not as well correlated as $R_b$ and $R_A$. Using standard
minimization techniques, we determine that a critical $\beta$ value of
$0.023$ has the smallest RMS distance between $R_b$ and $R_\beta$. We
note, however, that this may be a transitive effect; the mean value of
$\beta$ at $R_A$ is approximately $0.03$, with only minimal temporal
variation. Therefore, as $R_b$ is correlated with $R_A$ and
$\beta(R_A)\approx 0.03$, we expect the distance $R(\beta \approx
0.03)$ will be correlated with $R_b$.

\begin{figure}[ht]
\centering
\includegraphics[height=.95\linewidth,angle=90,viewport=60 90 545 715, clip=true]
{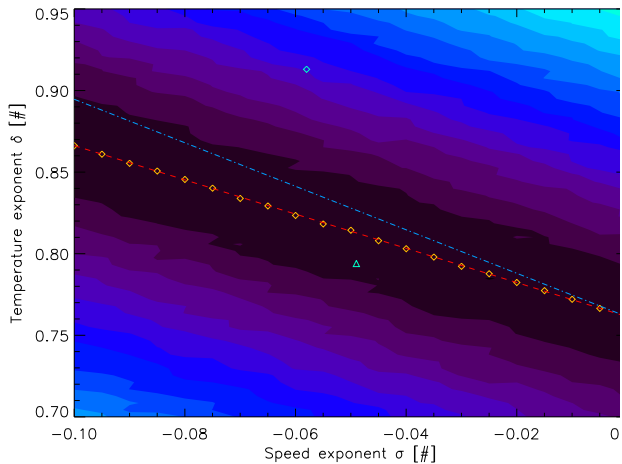}
\caption{The RMS difference between zone boundary $R_b$ and Alfv\'en
  point $R_A$, $\Delta R_{A,b}$, as a function of radial exponents $\sigma$ for speed and
  $\delta$ for temperature.  Published values for $\sigma$ in steady
  solar wind are generally between $-0.1$ and $0$.  Color
  contours indicate $1 R_\odot$ steps in the RMS difference, with the
  minimum values indicated by the orange diamonds. The red line indicates the
  best fit of relation between $\sigma$ and $\delta$ that passes
  through the RMS minimum.  The blue line is a prediction of this
  relation based on the degeneracy between two exponents in our
  equation for $A_c$. 
  This analysis results in the prediction that $\sigma$ and
  $\delta$ fall on the red line.  Points from \cite{Venzmer:2018} indicate values from two recent
  analyses of Helios radial trends.  
  }
\label{fig:rb_rms}
\end{figure}

We repeat the process of calculating $R_b$, $R_A$, $R_s$, and
$R_\beta$ for a range of radial power law exponents for proton
temperature and speed, $\delta \in [0.75,0.95]$ and $\sigma \in
[-0.1,0]$ compatible with radial trends extracted from Helios.
The Spearman rank correlation coefficients between $R_b$ and the three
critical surfaces have little variation due to the power law
exponents (not shown); for all values of $\delta$ and $\sigma$ considered, $R_A$
is by far the best correlated surface.  Further, we find that there is no
meaningful global minimum value for $\delta$ and $\sigma$ in the RMS
difference between $R_b$ and $R_A$, $\Delta R_{A,b}$ nor any global
maximum for their correlation coefficient.
Rather, there is a family
of solutions for which $\Delta R_{A,b}$ is minimized. 
Specifically, there is a unique value of $\delta$ for any $\sigma$ resulting in a minimum
$\Delta R_{A,b}(\sigma,\delta)$ of $\lesssim 5 R_\odot$, as illustrated in
Fig.~\ref{fig:rb_rms}.
The $\delta$ leading to the minimum $\Delta
R_{A,b}$ obeys the equation $\delta=0.813 -1.037\sigma$.  These
preferred solutions follow closely, but are not identical to the
$\delta = 2/3 - 4/3\sigma$ scaling suggested by the radial dependence
of the collisional age integral discussed in Paper I.
The minor disagreement between these trends is likely caused by uncertainty introduced by our model for radial variation in densities and Alfv\'en speeds.
\cite{Venzmer:2018} have produced limits on values for $\sigma$ and
$\delta$ based upon recent analysis of Helios radial trends which
bound the $\Delta R_{A,b}$ minimum extracted from the outer boundary
analysis. Using the mean value from that study, $\sigma =
-0.05$, we predict that $\delta = 0.85$.
 
\section{Discussion}

These results suggest that the outer boundary of the zone
of strong preferential ion heating $R_b$ is the Alfv\'en critical surface $R_A$, and that the zone and the Alfv\'en surface expand in lockstep as solar 
activity changes.  
This leads to the question of why any preferential heating mechanism
would be affected by a transition across this surface.  In the
expanding solar wind, some fraction of outward propagating wave-power
is reflected back toward the Sun due to large-scale gradients in
background quantities. Below $R_A$, these waves can travel all the way
back to the Sun and interact with outward propagating waves,
leading to wave-reflection driven
turbulence \citep{Matthaeus:1999,Perez:2013}. Above $R_A$, backward
propagating waves in the plasma frame are advected forward in the
Sun's reference frame.  The abundance of counter-propagating waves
below $R_A$ can dramatically enhance local preferential ion heating,
either due to wave-particle interactions\citep{Kasper:2013} or
alterations to the background
turbulence \citep{Velli:1989,Matthaeus:1999,Dmitruk:2001,Verdini:2007,Cranmer:2007,Chandran:2009,Verdini:2012}.
No theoretical predictions of dissipation have
suggested a sharp change in the preferential heating exactly at $R_A$.
Like crossing the event horizon of a black hole, there is no sudden
change experienced when crossing $R_A$, even as the plasma becomes
causally disconnected from the Sun.  While it is possible that there
is a discrete termination of preferential heating mechanisms, it
is more likely that the preferential heating gradually shuts off after
the plasma passes $R_A$.  Given that we find $\Delta R_{A,b}$ to
always be greater than $\approx 5 R_\odot$, this difference may serve as a estimate for the thickness over which the preferential heating ceases. 

Could there be some form of discontinuity or transition at the Alfv\'en point?  \citet{Weber:1967} proposed that the corona co-rotates with the Sun out to the Alfv\'en point, with a sudden drop in rotational speed as the solar wind Alfv\'en mach number exceeded unity.  Perhaps the diminished role of magnetic tension and the rotation of the magnetic field either alters the local turbulence or enhances reflection back towards the Sun.   Remote observations where the solar wind transitions character from striated to flocculated \citep{DeForest:2016} report distances of $44-88 R_\odot$, slightly beyond $R_A$, indicating that this transition region is not coterminous with $R_b$.  The Alfv\'en surface has been identified in numerical MHD
simulations\citep{Chhiber:2018} as the region where large-scale
magnetohydrodynamic turbulence first manifests, potentially changing
the mechanisms heating the plasma. 

We emphasize that we are only predicting that strong preferential ion
heating shuts off beyond $R_A$, not that all ion heating is terminated
outside of this zone. Weak preferential heating of minor ions may also continue outside of this zone, leading to no more than a ten percent difference between proton and alpha temperature, as reported for highly collisional solar wind at 1au\citep{Maruca:2013,Tracy:2016}; this temperature difference between species is also consistent with the temperature measurement error of the particle instruments\citep{Kasper:2006}.

\begin{figure}[ht]
\centering
\includegraphics[width=\linewidth,viewport=15 0 265 160, clip=true]
{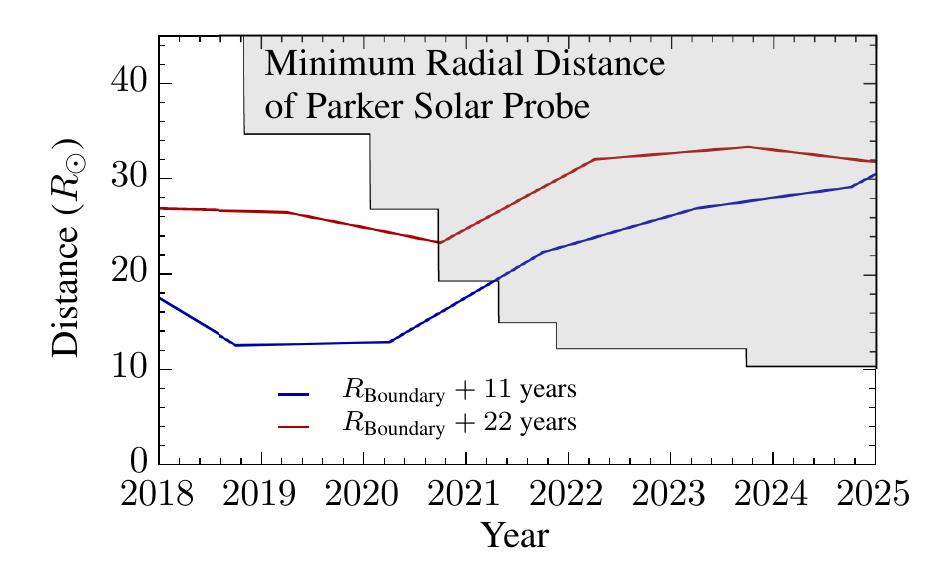}
\caption{The location of $R_b$ calculated from Wind observations, advanced 11 and 22 years in time in order to predict the motion of the outer boundary during the Parker Solar Probe (PSP) mission.  While PSP starts with a perihelion above $R_b$, we suggest the spacecraft and boundary will cross in late-2020, permitting the first direct observations of preferential ion heating.}
\label{fig:prediction}
\end{figure}

{
Parker Solar Probe (PSP), launched in August 2018, is the first spacecraft to enter the near-Sun environment, with an initial perihelion of $35 R_\odot$ in 2018 and final perihelia of $9.86 R_\odot$ starting in 2024 \citep{Fox:2016}.  The first scientific objective of PSP is to ``trace the flow of energy that heats and accelerates the solar corona and solar wind''.
By closing to within $10 R_\odot$ of the Sun's surface, PSP will have a high probability of observing non-thermal heating in action with its electromagnetic field\citep{Bale:2016} and ion and electron plasma \citep{Kasper:2016} instruments.  The explicit assumption has been that we are much more likely to observe this heating in action below the Alfv\'en point because we are in the magnetic atmosphere of the Sun, casually connected to the Sun, or simply because it may be easier to map the plasma to its sources that close to the Sun. 

With the results reported here we can make a specific prediction for how PSP may observe and reveal preferential ion heating in action, and the underlying physics, for the first time.  We have found that not only is there a zone of preferential ion heating surrounding the Sun that extends tens of $R_\odot$ from the Sun, but the outer boundary $R_b$ of this zone expands with solar activity, closely tracking the location of the Alfv\'en surface and likely intimately connected to changes in the nature of the plasma and waves across this surface.  In Fig.~\ref{fig:prediction}, we advance our calculation of $R_b$ using Wind observations by 11 and 22 years to project forward data from the last two solar cycles into the PSP mission timeframe.
Against this projection we plot the minimum distance of approach to the Sun by PSP, which steps closer to the Sun via six Venus gravity assists.  We find that at launch in solar minimum, PSP's perihelion is too high, and $R_b$ is too low, for PSP to enter the zone.  However, in late 2020 as PSP's perihelion lowers, the preferential heating zone and the Alfv\'en point will extend outwards and cross the trajectory of the spacecraft.  This prediction varies slightly when using $R_b$ from the relatively strong cycle 23 or the weaker cycle 24. See also \cite{Chhiber:2019} for a discussion of predictions from global MHD simulations for the Alfv\'en critical surface.

As $R_b$ rarely exceeds $35 R_\odot$, it is likely that no previous spacecraft has sampled this region of preferential heating.  As PSP approaches the zone, the derived distance to $R_b$ calculated using the method described in this work should decrease.  By comparing local measurements of the Alfv\'en speed and the solar wind bulk speed, we should be able to determine when PSP crosses the Alfv\'en point.  At that point we predict PSP will be able to detect if $\epsilon$ has reached its expected asymptotic value as well as signatures local heating processes in this region.
By doing so, PSP should be able to fulfill its first scientific objective of characterizing how the solar corona and solar wind are heated.
}

\section*{Acknowledgements}
All underlying data used for this analysis are archived and available for download at the NASA Space Physics Data Facility (\url{https://spdf.gsfc.nasa.gov/}).
We thank Tristan Weber for initial calculations of the Alfv\'en surface over time.  JCK is supported by Wind grant NNX14AR78G. KGK is supported by NASA HSR grant NNX16AM23G.

\bibliographystyle{apj.bst}



\end{document}